# The Physicist Inside the Ambiguous Room:
# an argument against the need of consciousness in the quantum mechanical measurement process

Carlo Roselli

The aim of this paper is to invalidate the hypothesis that consciousness is necessary in the quantum measurement process. In order to achieve this target, I propose a considerable modification of the *Schrödinger's cat and* the *Dead-Alive Physicist* thought experiments, called "PIAR", short for "Physicist Inside the Ambiguous Room". A *specific strategy has enabled me* to plan the experiment in such a way as to logically justify the inconsistency of the above hypothesis and to oblige its supporters to rely on an alternative interpretation of quantum mechanics in which a real world of phenomena exists independently of our conscious mind and where observers play no special role. Moreover, the description of the measurement apparatus will be complete, in the sense that the experiment, given that it includes also the experimenter, will begin and end exclusively within a sealed room. Hence, my analysis will provide a logical explanation of the relationship between the observer and the objects of her/his experimental observation; this and a few other implications will be discussed in the fifth section and in the conclusions.

Carlo Roselli, Rome (Italy), independent researcher; e-mail address: <beswick@tiscali.it>, <carloroselli39@gmail.com>; telephone +39-3335217868; +39-068543272

# 1 Introduction

This paper describes an alternative version of The Dead-Alive Physicist (DAP) experiment, called "The Physicist Inside the Ambiguous Room" (PIAR) and is aimed, as well as the DAP but through a considerably different approach,[1] at highlighting the inconsistency of the "idealistic" interpretation of quantum mechanics (QM).

The term "idealistic" is here used to refer to the (orthodox) Copenhagen [Niels Bohr (1885-1962)] view of atomic phenomena taken to the extreme. This view is based on two essential points: 1) a quantum system is in a state of genuine indeterminacy until it is measured; 2) the act of measurement forces the quantum system to adopt one of its potential states with a probability that can be calculated by means of the wave function (WF) which is appropriate for that system and for the measurement to which it is subject. Thus, according to the Copenhagen interpretation of QM, an elementary quantum phenomenon is not a phenomenon until such time as it is concluded by an *irreversible* measurement process and this would require some sort of explicit specification of the boundary that separates quantum from non quantum mechanical systems, given that the measurement process would be conceived as a non quantum mechanical phenomenon.

Unfortunately, the Copenhagen interpretation does not explain where and when a measurement process takes place. This omission gives rise to the so called "measurement problem", which weakens its claim to *completeness*. Indeed, this interpretation is not the only possible interpretation of QM that is subject to the measurement problem which, in more general terms, may be considered as that of defining a satisfactory transition process between micro-systems characterized by quantum state uncertainty and macro-systems obeying the deterministic laws of classical physics.

From the 1930's onwards, the measurement problem has been at the centre of a scientific-philosophical debate with the purpose of establishing when (or whether) the collapse of the WF occurs [6–27]. The debate on this issue has given rise to endless discussions among physicists and, so far, there has been a lack of consensus regarding which interpretation might be correct.

Furthermore, the Copenhagen interpretation of QM gives rise to some thought-provoking demonstrations, usually called "paradoxes", such as the *Einstein–Podolsky–Rosen* (EPR) [1], *Schrödinger's Cat* [2] and *Wigner's friend* experiments, which render questionable the theory's claim to completeness, unless one assumes that consciousness plays a fundamental role in the implementation of the quantum measurement process.

Eugene P. Wigner (1902-1995), following the books published in 1932 and 1955 by the mathematician John von Neumann [3-4] (1903-1957) and a little book published in 1939 by the physicists Fritz London and Edmond Bauer [5], developed an argument in favour of the consciousness assumption, leading to the thesis of the wave-function (WF) collapse at biological-mental level,[2] here more simply called "idealistic interpretation" of QM.

Starting from the orthodox view, the *idealistic interpretation* assumes that it is the observer's consciousness the fundamental factor which is able, in some unspecified and mysterious manner (Wigner refers to "a deus ex machina" [11, p. 188]) to collapse the quantum system down into one only of its possible states.

Here *consciousness* would not be playing a merely passive role in the measurement process, but would be the only factor capable of determining the transition from the ambiguous realm of potentials to the unequivocal realm of actual events. This is the kind of vision I mean when referring to the idealistic interpretation of QM.

---

[1] The PIAR experiment described in this paper, although largely similar to *The DAP experiment* (see ref. 35), involves an appreciable difference: while in the latter the observer's consciousness is present in one only branch of the wave function, in the former it is present in both branches.

[2] There are two main theses arguing that consciousness and quantum mechanical measurement are connected to each other: one thesis (von Neumann, London and Bauer, Wigner, Stapp; see refs 3-5 and 22 ) holds that the observer's consciousness causes the collapse of the wave function, thus claiming to complete the quantum-to-classical transition, while the other thesis (Penrose; Penrose and Hameroff; see refs 18-21 ) aims at demonstrating the opposite, i.e. that consciousness emerges from the so called "Orchestrated Objective Reduction".

By its very nature, the idealistic interpretation is difficult to evaluate, both in purely conceptual terms and, obviously, at the empirical-experimental level.

The objective of the thought experiment described hereafter is to demonstrate how the idealistic interpretation of QM, also known as "consciousness causes collapse (of the WF) hypothesis" (CCCH), is forced to conclusions incompatible with the assumption that consciousness[3] is necessary for providing a complete explanation of quantum measurement process.

## 2 The PIAR experiment

In this section I will propose an experiment, in which a male Physicist named "**P**" is inside an impenetrable room (Fig.1).

On the wall behind **P**, the apparatus L[4] is programmed to emit, at a precise time, a photon in a direction along which is placed a beam splitter (BS), that forms with it an angle of 45°. Two photo-detectors, $D_T$ and $D_R$, separated by an angle of 90° and both with 100% efficiency, are located beyond the BS; $D_T$ is fixed, along the direction of the transmitted photon, on the wall opposite to L; $D_R$ is fixed, along the direction of the reflected photon, on top of a box and is connected to a hooked hammer inside it; a Switch-On-Button (SOB) is installed under the hammer and, if pushed, it activates a buzzer (**B**).

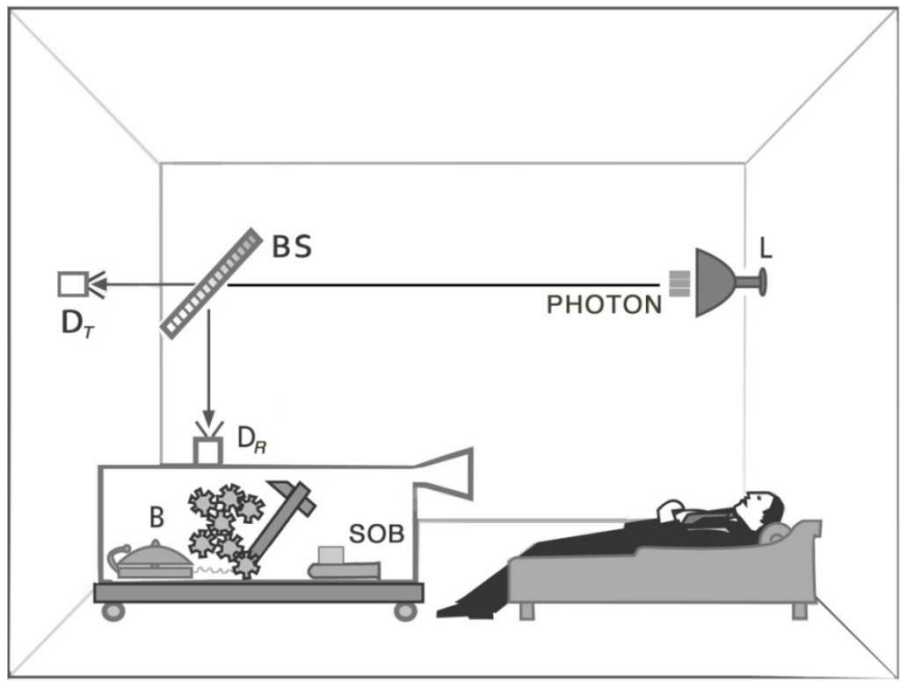

**Fig.1** The room, where **P**, under the **TCB** effect, will be unconscious from a few seconds after 1:00 PM to 2:00 PM.

---

[3] We don't know how consciousness works and do not have any idea of its nature. Nonetheless we are undoubtedly certain to possess it as the intimate and the most familiar of our experiences. In general, consciousness is defined as the faculty that allows a human subject to be aware of her/his self and of her/his mental activities, as well as the faculty to learn from the perception of external events to which these activities are directed. Leading contemporary scientists in the field have tried to lay the foundations for a science of consciousness, but none of them has yet been able to boast a promising theoretical approach. The Australian philosopher of the mind David J. Chalmers argues that, to open a window towards the understanding of consciousness, it would be required to solve the so called "difficult problem", consisting in finding a correlation between the functional mechanisms engendered by the neural activity of the brain and conscious experience, i.e. the phenomenon that allows the owner of that brain to feel specific effects in the first person.

[4] In such a mechanism a battery is supplying electric power.

The experiment is planned as follows:
1. L will emit a photon at 2.00 PM.
2. **P** has deliberately drugged himself one hour beforehand, at 1.00 PM, with a dose of a powerful narcotic, crucial for the experiment, called "**TCB**" (Temporary Consciousness Breaker) and 100% guaranteed to cut out conscious awareness for two hours and to prevent the later recall of events that occurred during the time of drug action; therefore, **P** will regain consciousness precisely at 3.00 PM; I say precisely at 3:00 PM, but I mean "precisely" when you can claim that **P** is no more under the **TCB** effect.
3. If $D_T$ registers, **B** will be inactive.
4. If $D_R$ registers, the hammer will be unhooked and, falling on the SOB, will activate **B** for over an hour.

Let us now briefly consider how the quantum theory describes the experiment: the photon is emitted from L at 2:00 PM, it collides with BS at 2:00 + $\Delta t_1$ PM (where $\Delta t_1$ is the travelling time of the photon from L to BS) and splits in two beams, one transmitted, *T*, moving along the direction of the detector $D_T$, and the other reflected, *R*, moving along the direction of the detector $D_R$, with a probability amplitude (in this example) of $1/\sqrt{2}$ for the photon to be received by each of them at 2:00 + $\Delta t_2$ PM (where $\Delta t_2$ is the travelling time of the photon WF from L to $D_T$ and $D_R$, both placed at the same distance from L).

According to the Copenhagen interpretation of QM, the WF $\psi$ is described as a linear superposition[5] of two states:

$$|\psi> = (|photon\ T> + |photon\ R>)/\sqrt{2} \qquad (1)$$

until the instant in which a measurement process takes place. This is the instant in which the WF collapses and only one of the two possible states becomes an actual outcome.[6]

If one assumes, as Wigner[9, p. 181 and p. 187], "the existence of an *influence* of consciousness on the physical world" and that "the measurement is not completed until a well-defined result enters our consciousness", that is until the WF collapses down into either one of its two component parts, then inside the room there is not one unique defined state as long as **P** is under the **TCB** effect, but rather *a linear superposition* of the two states described above which, while time is passing, is propagating along the whole macroscopic measurement system up to the scale of **P**'s brain. Then, the superposition will cease to be linear when it reaches **P**'s consciousness.

Consequently, there are two possibilities or chains of events, here called $E(_{T,\ R})$, which will travel according to the superposition principle until a certain instant of the experiment:

- $E_T$: *T* (part of the wave function transmitted), $D_T$ registers, **B** is inactive, **P** regains *consciousness* in the *silent room* (**SR**) precisely at 3:00 PM.
- $E_R$: *R* (part of the wave function reflected), $D_R$ registers, triggers the hammer, **B** is activated, **P** regains *consciousness* in the *buzzing room* (**BR**) precisely at 3:00 PM.

In this case, a *complete measurement apparatus* will be available, by this meaning that the experiment, since it includes the experimenter, will begin and end within the sealed room.

---

[5] A superposition of states can never be observed, since the system collapses to a single state at the instant that a measurement takes place.

[6] As well known, in QM experiments, the probabilities of measuring one or the other of two alternative outcomes are effectively the same calculated through formal procedure, according to the wave-packet reduction postulate based on the Born rule (introduced by Max Born); this rule, although being one of the most mysterious principles in quantum physics, is quite simple: it states that the probability of obtaining each of the possible outcomes is equal to the square of the corresponding amplitude. In our example the WF is the set of the two amplitudes described in equation (1), hence the probability is ½ for both the alternative outcomes.

## 3 Formal description of the PIAR experiment

All supporters of the CCCH may believe that any QM experiment, no matter whether applied to a cat or to a human being,[7] must give rise, in the end, to the same conclusions drawn by Wigner from his thought friend's experiment[8] and from his additional and stronger hypothesis [11, pp.185-196] concerning the role of consciousness in the quantum measurement process.

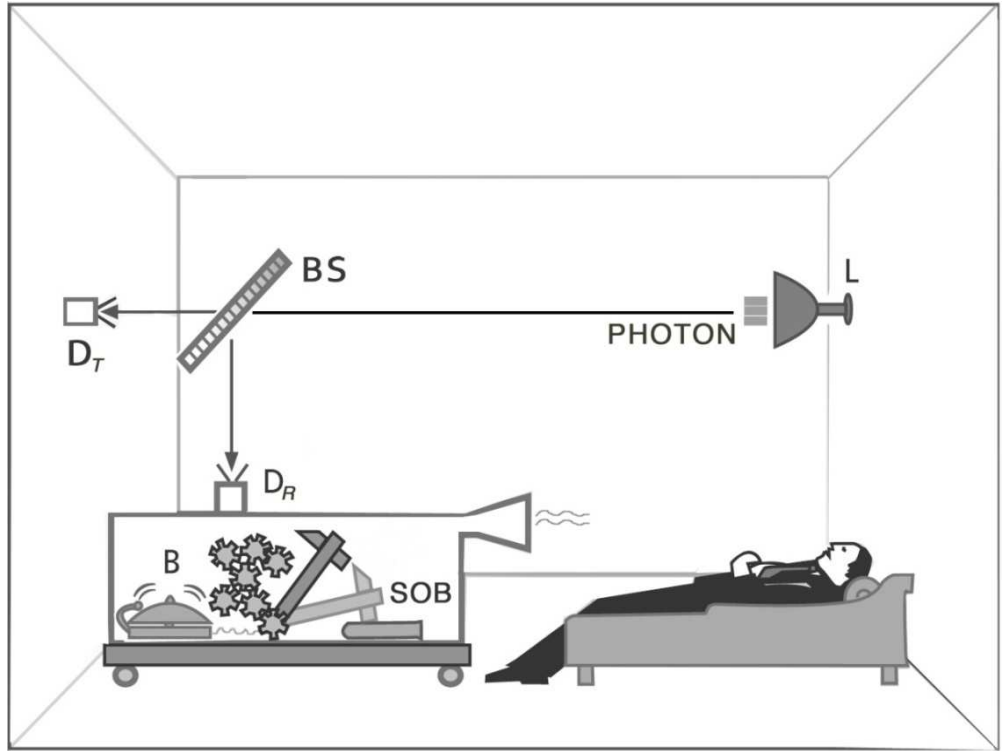

**Fig. 2** In the time interval between $2:00 + \Delta t_1$ and 3:00 PM, **P** is unconscious and the Copenhagen interpretation of QM would describe the system, in particular the buzzer **B**, in superposition of the states |**B** inactive> + |**B** active>.

In discussing the PIAR experiment described in the previous section, **P** will put himself in Wigner's place in order to verify whether or not there are the conditions for his consciousness to bring about the WF collapse (from now onwards also denoted with "WFC").

To this end, the orthodox interpretation of quantum mechanics would not preclude an observer situated outside the experimental room from describing the WF $\psi$, in the interval of time between $2:00 + \Delta t_1$ and 3:00 PM (Fig. 2), as follows:

---

[7] Observe that. in this experiment, QM is exceptionally applied to a human being, in spite of Wigner's conviction that it is not, applicable, as explained here below, at the end of footnote 8, point 3.

[8] Wigner's friend, here called "**F**", is a physicist left alone inside a laboratory with the task of checking *attentively* whether or not a detector has emitted a flash (has registered the arrival of a photon or not). Wigner is waiting outside and suspects that **F** (as well as all other human beings) may have weird perceptions and be in the superposition of macroscopically distinct states |**F** has perceived a flash> + |**F** has not perceived a flash>. Finally, Wigner enters the lab and asks **F** whether or not he perceived a flash. His reply (yes or no) should remove any doubt as to whether the wave-function collapse has occurred. However, Wigner will question whether it is acceptable or not to establish that the collapse into one only of the two possible alternatives is determined by his action (his request and reception of an unambiguous answer). He poses this question since his initial way of interpreting the state of the system gives rise to a rather embarrassing paradox, from which he has three possible ways of escape: 1) accept a relative form of *solipsism*, in the sense that he believes to be, among all living creatures, the only one who has unambiguous perceptions, 2) assume that QM is an incomplete theory, 3) assume that QM is not applicable to human beings; he refutes solipsism and, being a firm supporter of QM completeness, opts for the last solution, assuming that there are beings, at least human beings, endowed with *consciousness* that constitutes an ultimate reality and plays an active role in determining the measurement process by rules that are not susceptible to scientific description.

$$|\psi\rangle = (|\text{photon } T\rangle |D_T \text{ registers}\rangle |\text{the hammer hooked}\rangle |B \text{ inactive}\rangle + |\text{photon } R\rangle$$
$$|D_R \text{ registers}\rangle |\text{the hammer falls}\rangle | B \text{ active}\rangle)/\sqrt{2} \qquad (2)$$

Obviously, up to just before 3:00 PM, no information is available.

**3.1 Three remarks regarding the experiment**

    (i) - The expedient of the **TCB** has a fundamental function. In fact, supposing that it were not used, according to the CCCH, **P** would cause the WFC precisely as soon as he realizes either to perceive a buzzing or not.

    (ii) - Equation (2) describes a quantum superposition of two macroscopically distinguishable states in the interval of time between $2:00 + \Delta t_1$ and 3:00 PM. Note that, under the CCCH, the WFC requires a conscious observation, and in this experiment, as already mentioned at the end of section 2, **P** will be the only observer.

    (iii) - Human beings have access to their own internal states, perhaps similarly to cats or other animals, but, differently from these, they have the faculty of developing very sophisticated analytical thinking due to their cumulative culture.

**4 The PIAR experiment disproves the consistency of the CCCH**

    Up to just before 3:00 PM there is no conscious being in the game, and Wigner would say, consistently with the assumption of the **TCB** and his views about reduction, that the superposition is there (as depicted in figure 2): precisely at 3:00 PM, **P** becomes conscious and his consciousness causes the collapse of the WF into either the state '**SR**' or '**BR**'. But such a statement is *wrong*! Before explaining why, **P** considers it appropriate to make some reflections.

    Since the CCCH implies by definition a causal ordering[9] between two events, conscious observation and the WF collapse, **P** will examine whether, in the context of his experiment, there is a way to logically disprove the former as a causal agent of the latter**.** Hence, in order to achieve his aim, **P** will adopt a line of reasoning putting forward two crucial remarks.

    *First remark*: **P**, while planning his experiment, recognizes in it the occurrence of three distinct events rather than two: *the emergence of his consciousness*, *his perception/measurement of a well- defined outcome* and *the WF collapse*. He knows that he will regain consciousness precisely at 3:00 PM and that, at the same time, he will *perceive/measure* one of the two possible outcomes, **SR** or **BR**.[10] But **P** is sharp enough to understand, as will be explained later, that the WF collapse into either the state '**SR**' or the state '**BR**' cannot take place at 3:00 PM.

    *Second remark*: if you were a supporter of the CCCH and claimed that the WFC takes place when **P** regains consciousness, you would be easily misled to conclude that the *emergence* of consciousness (from now onwards also denoted with "**C"**) and the WF *collapse* into either the state '**SR**' or the state '**BR**' are two events representing, respectively, the *cause* and the *effect*, thus implying a well-defined causal ordering. But **P** would reject this conclusion, arguing that the CCCH will be never invalidated until the above two events can be thought of as causally ordered, formally as **C** -> **SR** or **C** -> **BR** and, implicitly, as **C** -> WFC.

    Then, **P** comprehends that the causal ordering **C** -> WFC does not fit the PIAR experiment, because this has been planned in such a way that either one of the two possible outcomes, **SR** or

---

[9] The question of causality is problematic, since it requires a distinction between the subjective and the objective aspects of this concept. Causality entails another (arguable) question called "cause and effect simultaneity", which has been discussed and investigated in depth by several philosophers, such as I. Kant, D. Hume, G.W. Leibniz and, recently, by Donald Gillies, Sylvain Bromberger et al.; for a detailed understanding see Buzzoni M.: *The Agency Theory of Causality, Anthropomorphism, and Simultaneity*, section 6, published online: 29 Jan 2015, https://doi.org/10.1080/02698595.2014.979668

[10] At 3:00 PM, **P** will be conscious and his consciousness, that cannot share the ambiguity of quantum world, may or may not perceive the buzzing; but in both cases, as it happens to Wigner's friend (who may or may not perceive a flash coming from the photon-detector placed near him), as according to the idealistic view of QM with perceptions, **P**'s consciousness collapses the wave function.

**BR**, will be perceived/measure by **P** precisely at the same time he regains consciousness (that is when the **TCB** effect is finished).

At last, **P** identifies in his experiment the decisive point that enables him to invalidate the CCCH. According to this hypothesis, the WF collapse is thought of as the immediate effect caused by the conscious perception/measurement of the outcome; but **P**, in his particular experiment, at 3:00 PM undergoes *concomitantly* two distinct events: the emergence of his consciousness and his perception/measurement of a well-defined outcome.

So, since **P** is certain that his consciousness cannot be in the game before he perceives/measures the *outcome*, a causal ordering, such as **C** -> **SR** or **C** -> **BR**, is logically inadmissible.

Finally, **P** concludes that the *outcome*, **SR** or **BR**, he will undergo together with the emergence of his consciousness, is not the WF collapse,[11] but should be a consequence of the collapse, reasonably occurring when either $D_T$ or $D_R$ registers the arrival of the photon at 2:00 + $\Delta t_2$.

Strictly speaking, in the PIAR experiment, each of the two events, the emergence of **P**'s consciousness and the WF collapse, happens on its own, so that they cannot in any way interact with each other.

This conclusion implies that the CCCH, as already substantiated above, is logically inconsistent despite a widely shared belief that it is not falsifiable (see for example J. Acacio de Barros and Gary Oas [28]).

**5 Conclusions**

If the above analysis is accepted as well-grounded, a supporter of the idealistic interpretation of QM should rely on an alternative interpretation, in which the role of the conscious observer is merely relegated to acknowledge the experimental results.

One can easily understand that the conclusion drawn at the end of the previous section has further implications, such as:

(**a**) - the concept of "collapse of the WF independently of consciousness" emerges from the *logical structure* of this thought experiment based on the **TCB** stratagem, since it allows to see in a new light the relationship between subject and object of observation, as shown in section 4;

(**b**) - if it were not conceivable an experiment capable of disproving the CCCH, this latter would still represent a possible and, for a few scientists, an even more suitable alternative to other interpretations of quantum mechanics.

(**c**) - in the realistic QM theories based on the collapse postulate, *the* boundary between quantum and classical systems should be *rescaled down*, reasonably to the transition point between the quantum system described in (1) and the initial (uncertain number of) atomic components of the photo-detector with which it interacts at 2:00 + $\Delta t_2$ PM;

(**d**) - in the collapse theories, *Schrödinger's cat* experiment can no longer be considered a paradox: before opening the box, the cat (as well as all the macroscopic measuring apparatuses inside the room until **P** is unconscious) is in a statistical mixture of states;

(**e**) - the falsification of the CCCH rules out also the hypothesis that the collapse of all the wave-functions involved in our Universe (according to the hypothesis shared by many scientists that consciousness is regarded as an emergent phenomenon) occurred when the first conscious human being appeared in it, thus avoiding to render the big-bang a senseless theory;

I think that the validity of the PIAR experiment is tenable with regard to one central hypothesis: the fact that, in certain controlled circumstances, *conscious perception* phenomenon, including self-awareness, could be *suspended* in a human subject. In other words, there could be an interval of time during which the subject is totally deprived of *self-awareness and the faculty of consciously perceiving* signals coming from the external surroundings. While this assertion may probably be open to doubt from a philosophical point of view, it appears sufficiently backed-up by common sense (and also by certain experiential data).

---

[11] Obviously, the time one acknowledges an event does not necessarily correspond to the effective time of its occurrence (in this very moment you could acknowledge something that happened long ago).

In synthesis, the starting point of this work is that the idealistic interpretation requires the superposition of macroscopically distinct states as well as the conscious perceptive faculty of the observer. This is necessary for consciousness to play a fundamental role in the WFC.

Nevertheless, it is possible to devise at least a thought experiment (e.g. the PIAR), which disproves the hypothesis that the WFC is caused by the observer's consciousness.

If this analysis is shared as logically compelling, then one is left with the immediate issue of what the best alternative to the idealistic interpretation should be, and clearly this is an entirely different (and daunting) problem.

However, I feel that the ordinary idea behind the PIAR experiment is that there are two ingredients given by the *wave function* and the observer's consciousness, which cannot in general be clearly separated, at least in such a way as to make the latter a causal agent in the collapse of the former. If this is true, then a fruitful way to tackle the measurement problem can only be one that treats the above two ingredients in a single coherent framework.

Recent advances in the quantum de-coherence and a re-examination of Everett's Many Worlds Interpretation suggest that such a framework could be constructed entirely within the boundaries of the theory itself; see, for instance, Roland Omnès [29], Maximilian Schlosshauer [30] and David Wallace [31], but clearly this is not the only route; see also Bernard d'Espagnat [32], the very recent works of Art Hobson [33-34] and Carlo Roselli, Bruno R. Stella [35].

Furthermore, the PIAR experiment, for being capable to disprove the consistency of the CCCH, could also represent a good reason for strengthening some of the actual quantum mechanical spontaneous localization models, where observers have no special role: I am referring to Ghirardi, Rimini and Weber theory (GRW), to Penrose and to Hameroff-Penrose interpretations, in which the WF is assumed to be as a physical reality and its collapse as an objective dynamical process, that in Penrose's approach is supposed to be induced by gravity.

I would like to close this paper quoting a sentence by Steven Weinberg [36, p. 124]:

"*I read a good deal of what had been written by physicists who had worried deeply about the foundations of quantum mechanics, but I felt some uneasiness at not being able to settle on any of their interpretations of quantum mechanics that seemed to me entirely satisfactory*".

**Acknowledgments**

My special gratitude goes to the late GianCarlo Ghirardi Professor Emeritus of Physics, Università di Trieste, Carlo Rovelli Professor of Physics, Université de Aix-Marseille, Art Hobson Professor Emeritus of Physics, University of Arkansas, Livio Triolo Professor (retired) of Mathematical Physics, Università Tor Vergata di Roma, Gianni Battimelli Professor (retired) of Physics, Università La Sapienza di Roma, Bruno Raffaele Stella Professor (retired) of Physics, Università Roma Tre and Enrico Marchetti, Professor of Economic Policy, Università degli Studi di Napoli Parthenope, for reading and commenting on the manuscript. Finally, I have to thank my wife Susan Jane Beswick for her scrupulous control of the English language of the text.-

**References**
1. Einstein A., Podolsky B. and Rosen N.: *Can Quantum Mechanical Description of Physical Reality Be Considered Complete?*, Phys. Rev. 47, 777-780 (1935).
2. Schrödinger, E.: *Die gegenwärtige Situation in der Quantenmechanik*, Naturwissenschaften 23, Heft 48, 807-812 (1935).
3. von Neumann, J.: *Mathematische Grundlagen der Quantenmechanik*, Julius Springer, Berlin (1932).

4. von Neumann, J.: *Mathematical Foundations of Quantum Mechanics*, Princeton University Press, Princeton NJ (1955).
5. London, F., Bauer, E.: *La Théorie de l'observation en méchanique quantique*, Hermann, Paris (1939).
6. Bohm, D.: *A suggested interpretation of the quantum theory in terms of "hidden" variables*, Phys. Rev. 85, 166–193 (1952).
7. Bohm, D.: *Quantum Theory*, Prentice-Hall, New York (1952).
8. Everett, H.: *'Relative State' Formulation of Quantum Mechanics*, Rev. Modern Phys. 29, 454-462 (1957).
9. Wigner, E.P.: *Remarks on the mind-body question.* In *The scientist speculates*, Good L. J., Editions W. Heinemann, London (1962).
10. Wigner, E.P.: *The problem of measurement*, Amer. J. Phys. 31, 6-15 (1963).
11. Wigner, E.P.: *Symmetries and Reflexions* , Indiana University Press, Indiana (1967).
12. Wheeler, J.A.: *Assessment of Everett's 'Relative State' Formulation of Quantum Theory*, Rev. Modern Phys. 29, 463-465 (1957).
13. DeWitt, B.S.: *The Many Worlds Interpretation of Quantum Mechanics*, Princeton Series in Physics, University Press, Princeton NJ (1973).
14. Zurek, W.H.: *Decoherence and the transition from quantum to classical*, Physics Today, 44, 36-44 (1991).
15. Albert, D.Z.: *Quantum Mechanics and Experience*, Harvard University Press, Cambridge, MA (1992).
16. Ghirardi, G.C., Rimini, A., Weber, T.: *Unified dynamics for microscopic and macroscopic systems*, Phys. Rev. D 34, 470 (1986).
17. Ghirardi, G.C., Grassi, R., Pearle, P.: *Relativistic dynamical reduction models: general framework and examples*, Found. Phys. 20, 1271–1316 (1990).
18. Penrose, R.: *The Emperor's New Mind*, Oxford University Press, Oxford (1989).
19. Penrose, R.: *Shadows Of The Mind*, Oxford University Press, Oxford (1994).
20. Penrose, R.: *The Large, the Small and the Human Mind*, Cambridge University Press, Cambridge MA (1997).
21. *Hameroff, S.* and *Penrose, R.: Orchestrated Reduction of Quantum Coherence in Brain microtubules: A model for consciousness. Math. Comput. Simulation, 40, 453-480 (1996).*
22. Stapp, H.P.: *Mind, Matter and Quantum Mechanics*, Springer. Berlin (1993).
23. Gell-Mann, M. and Hartle, J.: *Strong decoherence*. In: Feng, D.H., Hu, B.L. (eds.), *Quantum classical correspondence*. The 4th Drexel Symposium on Quantum Nonintegrability, pp. 3–35, International Press, Cambridge MA (1997).
24. Rovelli, C.: *Relational Quantum Mechanics*, Internat. J. Theoret. Phys. 35, 8, 1637-1678 (1996).
25. Tegmark, M.: [*The Interpretation of Quantum Mechanics: Many Worlds or Many Words?*](#) Fortschr. Phys. 46, 855-862 (1998).
26. Haroche, S.: *Entanglement, decoherence and the quantum/classical boundary*, Phys. Today 51, 36–42(1998).
27. Bell, J.S.: *Speakable and Unspeakable in Quantum Mechanics:* Collected Papers on Quantum Philosophy, Cambridge University Press, Cambridge MA (2004).
28. de Barros, J. A., Oas, G.: *Can we Falsify the Consciousness-Causes-Collapse Hypothesis in Quantum Mechanics?* Found.Phys. 47, 1294-1308 (2017).

29. Omnès, R.: *Converging realities: Toward a common Philosophy of Physics and Mathematics*, Princeton University Press. Princeton (2004).
30. Schlosshauer, M.: *Decoherence and the Quantum-to-Classical Transition*, Springer, Berlin (2007).
31. Wallace, D.: *The Emergent Multiverse: Quantum Theory according to the Everett Interpretation*, Oxford University Press, Oxford (2012).
32. d'Espagnat, B.: *Quantum Physics and Reality*, Found. Phys. 41, 1703-1716 (2011).
33. Hobson, A.: *Review and suggested resolution of the problem of Schrodinger's cat*, Contemp. Phys. 59, 16-30 (2018)
34. Hobson, A.: *Entanglement and the measurement problem*, arXiv:2002, [quant-ph] (25 Feb 2020).
35. Roselli, C. and Stella, B. R.: *The Dead-Alive Physicist experiment: a case study against the hypothesis that consciousness causes the wave function collapse in the quantum mechanical measurement process*, Found. Phys. 51, Issue 21, (2021)
36. Weinberg, S.*: The Trouble with Quantum Mechanics in Third Thoughts*, Harvard University Press, Cambridge, Massachusetts (2018)**.-**